\documentclass[aps,prb,reprint,showpacs,floatfix,superscriptaddress]{revtex4-1}
\usepackage{amsmath,amssymb,amsfonts,graphicx,placeins,float,bm}
\usepackage{hhline,enumerate}
\usepackage{hyperref}
\usepackage{blindtext}
\usepackage{xcolor}
\usepackage{printlen}
\newcommand{\bs}{\boldsymbol}

\begin{document}
	\title{Spin-polarized currents and noise in NS junctions with Yu-Shiba-Rusinov impurities}

	\author{Daniel Persson}
	\email{daniel.j.persson@chalmers.se}
	\affiliation{Department of Microtechnology and Nanoscience -- MC2, Chalmers University of Technology, SE-412 96 G\"{o}teborg, Sweden}

	\author{Oleksii Shevtsov}
	\email{oleksii.shevtsov@northwestern.edu}
	\affiliation{Department of Microtechnology and Nanoscience -- MC2, Chalmers University of Technology, SE-412 96 G\"{o}teborg, Sweden}
	\affiliation{Department of Physics and Astronomy, Northwestern University, Evanston, IL 60208 USA}
	
	\author{Tomas L\"{o}fwander}
	\email{tomas.lofwander@chalmers.se}
	\affiliation{Department of Microtechnology and Nanoscience -- MC2, Chalmers University of Technology, SE-412 96 G\"{o}teborg, Sweden}

	\author{Mikael Fogelstr\"{o}m}
	\email{mikael.fogelstrom@chalmers.se}
	\affiliation{Department of Microtechnology and Nanoscience -- MC2, Chalmers University of Technology, SE-412 96 G\"{o}teborg, Sweden}

	\date{\today}


	\begin{abstract}
	Conventional superconductors disordered by magnetic impurities demonstrate physical properties drastically different from their pristine counterparts. In our previous work [Phys.~Rev.~B~\textbf{92}, 245430 (2015)] we explored spectral and thermodynamic properties of such systems for two extreme cases: completely random and ferromagnetically aligned impurity magnetic moments. Here we consider transport properties of these systems, and show that they have a potential to be used in superconducting spintronic devices. Each magnetic impurity contributes a Yu-Shiba-Rusinov (YSR) bound state to the spectrum, residing at sub-gap energies. Provided the YSR states form metallic bands, we demonstrate that the tunneling current carried by these states can be highly spin-polarized when the impurities are ferromagnetically ordered. The spin polarization can be switched by simply tuning the bias voltage. Moreover, even when the impurity spins are completely uncorrelated, one can still achieve almost $100\%$ spin polarization of the current, if the tunnel interface is spin-active. We compute electric current and noise, varying parameters of the interface between tunneling and fully transparent regimes, and analyze the relative role of single-particle and Andreev reflection processes.
	\end{abstract} 

	\maketitle

\section{Introduction}
  The superconducting condensate in conventional superconductors is formed of spin-singlet Cooper pairs, i.e. correlated pairs of electrons with opposite spins. Therefore, spin-dependent scattering induced by the presence of magnetic impurities or proximity of a ferromagnet can be detrimental for superconductivity. Indeed, it was shown by Abrikosov and Gor'kov \cite{Abrikosov1960} that magnetic impurities lower the superconducting transition temperature and can even lead to gapless superconductivity. However, the effect of spin-dependent scattering is not always negative. For example, spin-triplet correlations between  equal-spin electrons can be induced in hybrid structures consisting of ferromagnets and conventional superconductors \cite{Bergeret2005,Buzdin2005,Eschrig2007}. This feature makes such systems promising for applications in superconducting spintronic devices \cite{Eschrig2011,Linder2015,Eschrig2015}. On the other hand, specially tailored arrangements of magnetic impurities, such as linear chains for example, are able to host the elusive Majorana excitations at their boundaries \cite{Nadj-Perge2013,Pientka2013,KlinovajaPRL2013,Nadj-Perge2014,Peng2015,SacramentoJPCM2015}.
  
  In the late 1960s it was shown that classical spins in superconductors induce bound states at sub-gap energies \cite{Yu1965,Shiba1968,Rusinov1969}, now termed Yu-Shiba-Rusinov (YSR) states. Experimental control of individual impurities has reached the point when it is possible to access the local properties of the system in vicinity of each impurity \cite{Ruby2015,Menard2015}. For example, in Ref.~[\onlinecite{Ruby2015}] Ruby \textit{et al.} studied a superconductor with magnetic adatoms deposited on its surface. They were able to investigate microscopic tunneling processes between an STM tip and a YSR state of a single impurity by varying the distance between the tip and the sample. They observed different transport regimes governed by the relative role of single-particle and Andreev reflection processes. If we imagine a homogeneous distribution of such impurities, then the YSR states may overlap and form metallic bands, supporting electric current at sub-gap voltages. It is important to investigate the nature of elementary charge carriers in this case, especially when the impurity subsystem gets correlated (ferromagnetically aligned, for example).
  
  In this work we investigate transport properties of conventional $s$-wave spin-singlet superconductors with magnetic impurities. Assuming that an electric current is injected in such a system via a normal-metal probe (STM tip, for example), we perform an extensive analysis of both current and noise. Computing the differential Fano factor \cite{Blanter2000} allows us to decipher the relative role of single-particle and two-particle (Andreev reflection) processes in the tunneling current. When the impurities are ferromagnetically aligned, the electric current supported by the YSR bands can reach $100\%$ spin polarization in the tunneling limit. At the same time, in the appropriate parameter regime, the current polarization can be switched between spin-up and spin-down by simply tuning the applied bias. Finally, if we allow the tunnel interface to be spin-active \cite{Tokuyasu1988,Fogelstrom2000,Zhao2004}, one can still observe completely spin-polarized current via the YSR bands, even when the magnetic impurities in the superconductor are completely uncorrelated.

  The paper is organized as follows. In Sec. \ref{Sec_Model} we discuss the theoretical model and introduce the framework used to solve the problem. In Sec. \ref{Sec_Results} we present the main results, considering three different cases: (i) tunneling limit (low transparency), (ii) high transparency, and (iii) spin-active interface and the inverse proximity effect. For each case we allow for the magnetic impurity spins to be ferromagnetically ordered or randomly oriented. Finally, in Sec. \ref{Sec_Concl} we discuss the obtained results and conclude.
	
\section{Theoretical Model}\label{Sec_Model}
\subsection{Quasiclassical Green's function}
	We use the quasiclassical theory of superconductivity, which is an extension of Fermi-liquid theory to encompass superconducting \cite{Eilenberger1968,Larkin1969} and superfluid phenomena \cite{Serene1983}. It is based on separations of energy and length scales relevant in normal and superfluid phases, namely $E_F\gg\Delta$ or $\lambda_F\ll\xi_0$. Here, $E_F$ and $\lambda_F$ are the Fermi energy and wavelength, while $\Delta$ and $\xi_0=\hbar v_F/2\pi k_BT_c$ are the order parameter and coherence length in a superconductor ($T_c$ is the critical temperature). The central object of the theory is the single-particle Green's function. Starting from the full Gor'kov equations for a mean-field BCS Hamiltonian, and keeping only the coherence length scale variations in the system, we end up with the quasiclassical Eilenberger equation \cite{Eilenberger1968}
	\begin{equation}
		\left[\varepsilon\hat{\tau}_3\check{1}-\check{h},\check{g}\right]+i\hbar \mathbf{v}_F \cdot\nabla \check{g}=\check{0},
		\label{Eilenberger}
	\end{equation}	
	together with the normalization condition
	\begin{equation}
		\check{g}^2=-\pi^2\check{1}.
		\label{Eilenberger_norm}
	\end{equation}
  The quasiclassical Green's function $\check{g}(\varepsilon,\mathbf{p}_F,\mathbf{r})$ and the self-energy $\check{h}(\varepsilon,\mathbf{p}_F,\mathbf{r})$ are, in general, matrices in Keldysh (denoted by check) $\times$ Nambu (denoted by hat) $\times$ spin space,
  \begin{align}
  \label{gKeldMatr}
  \check{\chi}=
  \begin{pmatrix}
  \hat{\chi}^{R} & \hat{\chi}^{K} \\
  0 & \hat{\chi}^{A} 
  \end{pmatrix},\;\; \chi = \left\{g,h\right\}.
  \end{align} 
  Here, $\varepsilon$ is the energy, $\mathbf{p}_F$ is a point on the Fermi surface, and $\mathbf{r}$ is the spatial coordinate. For brevity, we will avoid writing the arguments explicitly, except where it is needed. We introduce two sets of Pauli matrices, $\hat{\boldsymbol{\tau}} = \left(\hat{\tau}_1,\hat{\tau}_2,\hat{\tau}_3\right)$ and $\boldsymbol{\sigma} = \left(\sigma_x,\sigma_y,\sigma_z\right)$, to resolve the matrix structure in Nambu and spin spaces, respectively.
\subsubsection{Riccati parametrization}
	We employ the so-called Riccati parametrization \cite{Schopohl1998,Eschrig2000,Eschrig2009}, which is a convenient way of solving the Eilenberger equation, Eq.~(\ref{Eilenberger}). For the retarded and advanced components of the Green's function it reads,
	\begin{equation}
		\begin{aligned}
			\hat{g}^{R,A}&=\mp2\pi i 
			\begin{pmatrix}
				\mathcal{G} & \mathcal{F} \\
				-\tilde{\mathcal{F}} & -\tilde{\mathcal{G}}
			\end{pmatrix}^{R,A}
			\pm i\pi\hat{\tau}_3,\\
			&\mathcal{G}=\left(1-\gamma\tilde{\gamma}\right)^{-1},\,\,\mathcal{F}=\mathcal{G}\gamma,
			\label{Riccati_para}
		\end{aligned}
	\end{equation}
	where the ``tilde''-operation expresses particle-hole conjugation, defined as 
	  \begin{equation}
	  \tilde{A}(\varepsilon,\mathbf{p}_F,\mathbf{r})=A(-\varepsilon^*,-\mathbf{p}_F,\mathbf{r})^*.
	  \label{tilde_def}
	  \end{equation}
	One should understand $\varepsilon$ here as a real quantity for the Keldysh component, and having non-zero positive (negative) imaginary part for the retarded (advanced) one.
	The parametrization defined in Eq.~(\ref{Riccati_para}) automatically satisfies Eq.~(\ref{Eilenberger_norm}) and transforms Eq.~\eqref{Eilenberger} into a set of differential equations for the two coherence functions 
	$\gamma(\varepsilon,\mathbf{p}_F,\mathbf{r})$ and $\tilde{\gamma}(\varepsilon,\mathbf{p}_F,\mathbf{r})$,
	\begin{equation}
		\begin{aligned}
			&\left(i\hbar\mathbf{v}_F \cdot\nabla+2\varepsilon\right)\gamma^{R,A}\!=\!
			\left[\gamma\tilde{\Delta}\gamma+\Sigma\gamma-\gamma\tilde{\Sigma}-\Delta\right]^{R,A},\\
			&\left(i\hbar\mathbf{v}_F \cdot\nabla-2\varepsilon\right)\tilde{\gamma}^{R,A}\!=\!
			\left[\tilde{\gamma}\Delta\tilde{\gamma}+\tilde{\Sigma}\tilde{\gamma}-\tilde{\gamma}\Sigma-\tilde{\Delta}\right]^{R,A},
		\end{aligned}
		\label{Riccati}
	\end{equation}
	where we have used the following representation of the self-energies in Nambu space,
	\begin{align}
	\hat{h}^{R,A}=
	\begin{pmatrix}
	\Sigma & \Delta \\
	\tilde{\Delta} & \tilde{\Sigma}
	\end{pmatrix}^{\!\!R,A},\;\;
	\hat{h}^{K}=
	\begin{pmatrix}
	\Sigma & \Delta \\
	-\tilde{\Delta} & -\tilde{\Sigma}
	\end{pmatrix}^{\!\!K}.
	\label{SEs_par}
	\end{align}
	We note that the coherence functions and the self-energies here still have a $2\times2$ matrix structure in spin-space. 
	The advanced coherence functions are related to the retarded ones via
	\begin{equation}
		\gamma^A=\left[\tilde{\gamma}^R\right]^\dagger,
		\label{RetAdv}
	\end{equation}
	meaning that it is sufficient to solve for retarded quantities only. Finally, since we are interested in transport properties, we also need to know the Keldysh component of the Green's function, which is parametrized by means of the two distribution functions $x(\varepsilon,\mathbf{p}_F,\mathbf{r})$ and $\tilde{x}(\varepsilon,\mathbf{p}_F,\mathbf{r})$ as
  \begin{equation}
    \hat{g}^K=-2\pi i
    \begin{pmatrix}
      \mathcal{G} & \mathcal{F}\\
      -\mathcal{\tilde{F}} & -\mathcal{\tilde{G}}
    \end{pmatrix}^R
    \begin{pmatrix}
      x & 0\\
      0 & \tilde{x}
    \end{pmatrix}
    \begin{pmatrix}
      \mathcal{G} & \mathcal{F}\\
      -\mathcal{\tilde{F}} & -\mathcal{\tilde{G}}
    \end{pmatrix}^A.
    \label{GreensK}
  \end{equation}
  This leads to the two additional differential equations
	\begin{equation}
    \begin{aligned}
      i\hbar\mathbf{v}_F\cdot\nabla x-\left[\gamma\tilde{\Delta}+\Sigma\right]^R x-x \left[\Delta\tilde{\gamma}-\Sigma\right]^A&=\\
      -\gamma^R\tilde{\Sigma}^K\tilde{\gamma}^A+\Delta^K\tilde{\gamma}^A+\gamma^R\tilde{\Delta}^K-\Sigma^K,\\
      i\hbar\mathbf{v}_F\cdot\nabla\tilde{x}-\left[\tilde{\gamma}\Delta+\tilde{\Sigma}\right]^R \tilde{x}
      -\tilde{x}\left[\tilde{\Delta}\gamma-\tilde{\Sigma}\right]^A&=\\
      -\tilde{\gamma}^R\Sigma^K\gamma^A+\tilde{\Delta}^K\gamma^A+\tilde{\gamma}^R\Delta^K-\tilde{\Sigma}^K.
    \end{aligned}
    \label{RiccatiK}
  \end{equation}
  
  As was mentioned above, we are going to consider the two extreme configurations of magnetic impurities spins, namely randomly oriented and ferromagnetically aligned. In this case our problem has at most one given spin quantization axis. Based on this, we can write down quite generally the following expression for the coherence functions,
  \begin{equation}
    \gamma^{R,A}=
    \begin{pmatrix}
      \gamma_\uparrow & 0\\
      0 & \gamma_\downarrow
    \end{pmatrix}^{\!\!R,A}
    i \sigma_y,\;\;
    \tilde{\gamma}^{R,A}=
    \begin{pmatrix}
      \tilde{\gamma}_\uparrow & 0\\
      0 & \tilde{\gamma}_\downarrow
    \end{pmatrix}^{\!\!R,A}i \sigma_y.
    \label{gamma_diag}
  \end{equation}
  This allows to split the problem into two sub-problems for different spin bands. Obviously, for the case of randomly oriented spins, $\gamma_\uparrow = \gamma_\downarrow$.
  \subsubsection{Self energies}
  In order to solve Eqs. \eqref{Riccati} and \eqref{RiccatiK} we also have to formulate self-consistency equations for the self-energies and superconducting order parameter. The self-energy matrix $\hat{h}$, see Eq.~(\ref{SEs_par}), contains both impurity contributions and the order parameter. For our purposes it is enough to consider only retarded and advanced self-energies. We adopt the self-energies obtained in our recent work, see Ref.~[\onlinecite{Persson2015}], where we have thoroughly studied thermodynamic properties of the model considered here. For a self-contained presentation, we give a short recap of the basic equations and parameters below.
  
  The order parameter of a spin-singlet $s$-wave superconductor has the form $\Delta^R_0(\mathbf{r}) = \Delta_0(\mathbf{r})i\sigma_y$, where
  \begin{align}
  \Delta_0(\mathbf{r})&=\frac{\lambda N_{F}}{16\pi i}\int_{-\varepsilon_c}^{\varepsilon_c} d\varepsilon\int\frac{d\Omega_{\mathbf{p}_{F}}}{4\pi}\notag\\
  &\times\mathrm{Tr}\left[i\sigma_y(\hat{\tau}_1-i\hat{\tau}_2)\hat{g}^{\mathrm{K}}(\varepsilon,\mathbf{p}_{F},\mathbf{r})\right].
  \label{OP_gen}
  \end{align}
  Here, $\lambda < 0$ is the electron-phonon coupling constant, $N_F$ is the normal-state density of states per spin at the Fermi level, and $\varepsilon_c$ is the high-energy cut-off of the order of the Debye frequency. Magnetic impurities are treated within the non-crossing t-matrix approximation \cite{Rammer1998}, and the impurity self-energy is given by the single-impurity $t$-matrix multiplied by the density of impurities $n$,
	\begin{equation}
	\hat \Sigma_{\mathrm{imp}}(\varepsilon,\mathbf{p}_F)=n \hat t_{\mathrm{imp}}(\varepsilon,\mathbf{p}_F,\mathbf{p}_F).
	\end{equation}
	The matrix $\hat t_{imp}$ satisfies
	\begin{multline}
		\hat{t}_{\mathrm{imp}}(\varepsilon,\mathbf{p}_{F},\mathbf{p}_{F}^{\prime}) = \hat{v}(\mathbf{p}_{F},\mathbf{p}_{F}^{\prime})\\
		+N_F\int\frac{d\Omega_{\mathbf{p}_{F}^{\prime\prime}}}{4\pi}\hat{v}(\mathbf{p}_{F},\mathbf{p}_{F}^{\prime\prime})
		\hat{g}(\varepsilon,\mathbf{p}_{F}^{\prime\prime})\hat{t}_{\mathrm{imp}}(\varepsilon,\mathbf{p}_{F}^{\prime\prime},\mathbf{p}_{F}^{\prime}).
		\label{t_matr_eqn}
	\end{multline}
	Here, $\hat{v}(\mathbf{p}_{F},\mathbf{p}_{F}^{\prime})$ is the matrix element of the impurity potential between the quasi-particle states with 
	momenta $\mathbf{p}_{F}$ and $\mathbf{p}_{F}^{\prime}$ on the Fermi surface (computed in the normal state of the system).
	Below we consider only s-wave scattering off impurities, i.e. $\hat{v}(\mathbf{p}_{F},\mathbf{p}_{F}^{\prime})$ is independent of momenta.	Then, it can be written as
	\begin{equation}
		\hat{v} = 
		\begin{pmatrix}
			v & 0\\
			0 & v^{\ast}
		\end{pmatrix},
		\;\;v = v_0+\alpha v_{\mathrm{S}}\mathbf{m}\cdot\bs{\sigma},
	\end{equation}
	where $v_0$ parameterizes the scalar part, and $v_{\mathrm{S}}$ the exchange part of the scattering.  
	
	For the case of unpolarized magnetic impurities, besides averaging over impurity positions, one also has to average Eq.~(\ref{t_matr_eqn}) over the magnetic moment directions, obtaining the self-energy 
	\begin{equation}
		\hat{\Sigma}_{\mathrm{imp}}(\varepsilon) = n\langle\hat{t}_{\mathrm{imp}}(\varepsilon)\rangle_{\mathrm{spin\:dir.}}.
	\end{equation}
	For the case of ferromagnetically ordered magnetic impurities we can choose the coordinate system in spin space such as 
	$\mathbf{m}_j \equiv \mathbf{m} = (0,0,1)$. 
	Since in this case, apart from the local scattering by the impurities, we also have a background magnetic field in the system, 
  the impurity self-energy consists of two parts \cite{Persson2015}
	\begin{equation}
		\hat{\Sigma}_{\mathrm{imp}}(\varepsilon) = \beta n v_{\mathrm{S}}\sigma_{z}\hat{1}+n\hat{t}_{\mathrm{imp}}(\varepsilon),
		\label{SE_aligned}
	\end{equation}
	where the first term has a form of Zeeman interaction, while the second one is obtained by solving Eq.~(\ref{t_matr_eqn}).  
	
  We define a set of parameters for our impurity model: the scattering rate $\Gamma=n/\pi N_F$, the dimensionless scalar $u_0=\pi N_F v_0$, and the exchange $u_{\mathrm{S}}=\pi N_F v_{\mathrm{S}}$ parts of the impurity potential.
  The two additional parameters $\alpha$ and $\beta$ are needed to describe the spin-polarized case. Coupling via tunneling of itinerant electrons on and off the impurity site is given by $\alpha$, where the sign of $\alpha$ discriminates between (anti-)ferromagnetic exchange coupling $\alpha < 0 ( > 0)$, see Ref.~[\onlinecite{Persson2015}]. The parameter $\beta\sim 1$ is a dimensionless fitting parameter, related to the geometrical structure factor of the actual impurity distribution in space. 
  For the numerical results presented below we assume that $\beta=1$ and $|\alpha|=0.1$. Dependence of the system characteristics on the scalar part of the impurity potential, $u_0$, is weak. In Ref.~[\onlinecite{Okabe1983}] it was demonstrated that for an isotropic order parameter (as we have here) the scalar part only enters the theory through the position of the YSR bound state. Introducing an effective exchange scattering amplitude, $\left\{u_0,u_{\mathrm{S}}\right\}\rightarrow u_{\mathrm{S}}^{\mathrm{eff}}$, this can be completely accounted for. Therefore, we take $u_0=0$ in all our results.
  \begin{figure}[b]
  	\centering
  	  \includegraphics[width=\columnwidth]{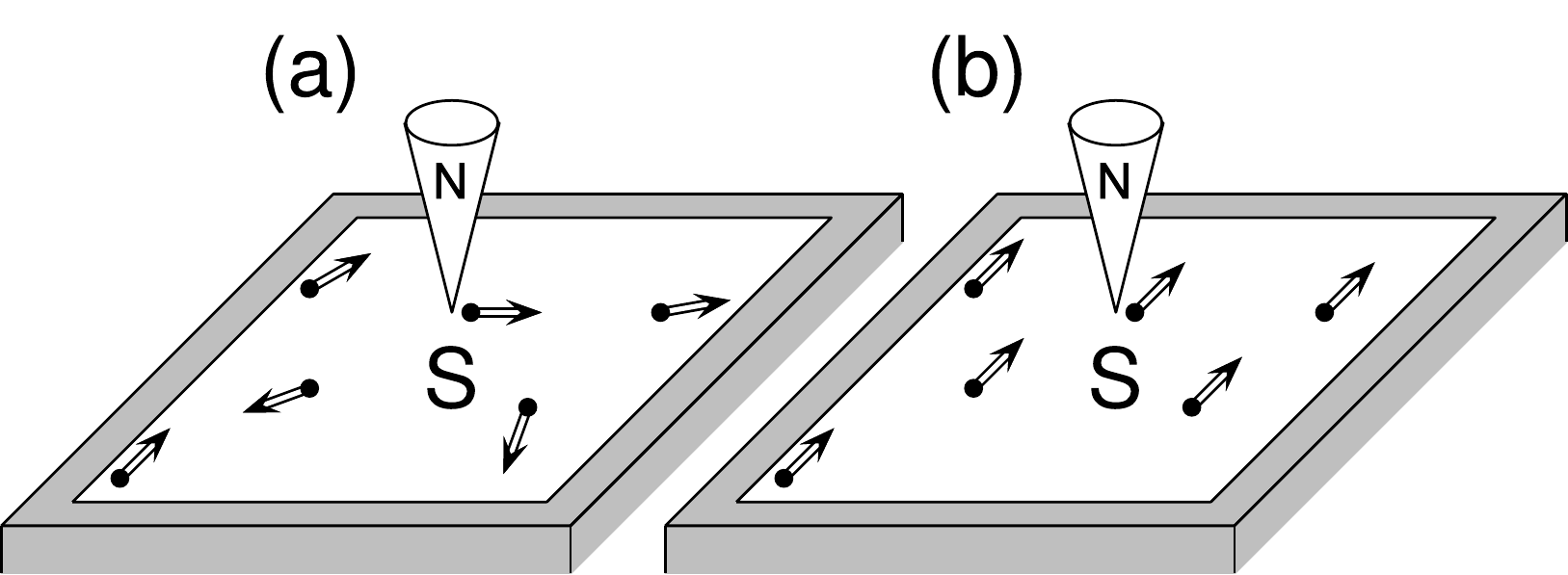}
  	  \caption{\label{Fig_junc}
  	    Superconducting thin film (S) is deposited on a substrate and magnetic impurities are homogeneously distributed within the sample. The magnetic moments of the impurities are either completely unpolarized (a), or ferromagnetically ordered (b) in the plane of the film. The film is contacted by a normal-metal probe (N) for transport measurements.}
  	\end{figure}
  	
  \begin{table}[t]
  	\centering
  	\caption{\label{scatter_amplitudes2}
  		All possible scattering amplitudes at an NS-junction, written for excitations originating in N and S. We define,
  		$D^{R} = (1-\mathcal{S}_\mathcal{R}\gamma^{R}\tilde{\mathcal{S}}_\mathcal{R}\tilde{\gamma}^{R})^{-1}$ and $R^{R} = S_\mathcal{R}-S_\mathcal{D}\gamma^R\tilde{\mathcal{S}}_\mathcal{R}\tilde{\gamma}^RD^RS_\mathcal{D}$.
  	}
  	\renewcommand{\tabcolsep}{2.4em}
  	\renewcommand{\arraystretch}{1.5}
  	\begin{tabular}{l|l}
  		\hline\hline
  		Incident from N: & Incident from S:\\
  		$r^R_{ee}=R^R$ & $\overline{r}^R_{ee}=-D^R\mathcal{S}_\mathcal{R}$\\
  		$r^R_{eh}=\mathcal{S}_\mathcal{D}\gamma^R\tilde{D}^R\tilde{\mathcal{S}}_\mathcal{D}$ & $\overline{r}^R_{eh}=\mathcal{S}_\mathcal{R}\gamma^R\tilde{D}^R\tilde{\mathcal{S}}_\mathcal{R}$\\
  		$r^R_{he}=\tilde{\mathcal{S}}_\mathcal{D}\tilde{\gamma}^RD^R\mathcal{S}_\mathcal{D}$ &  $\overline{r}^R_{he}=\tilde{\mathcal{S}}_\mathcal{R}\tilde{\gamma}^RD^R\mathcal{S}_\mathcal{R}$\\
  		$r^R_{hh}=\tilde{R}^R$ & $\overline{r}^R_{hh}=-\tilde{D}^R\tilde{\mathcal{S}}_\mathcal{R}$\\
  		\hline
  		$t^R_{ee}=D^R\mathcal{S}_\mathcal{D}$ & $\overline{t}^R_{ee}=D^R\mathcal{S}_\mathcal{D}$\\
  		$t^R_{eh}=-\mathcal{S}_\mathcal{R}\gamma^R\tilde{D}^R\tilde{\mathcal{S}}_\mathcal{D}$ & $\overline{t}^R_{eh}=-\mathcal{S}_\mathcal{D}\gamma^R\tilde{D}^R\tilde{\mathcal{S}}_\mathcal{R}$\\
  		$t^R_{he}=-\tilde{\mathcal{S}}_\mathcal{R}\tilde{\gamma}^RD^R\mathcal{S}_\mathcal{D}$ & $\overline{t}^R_{he}=-\tilde{\mathcal{S}}_\mathcal{D}\tilde{\gamma}^RD^R\mathcal{S}_\mathcal{R}$\\
  		$t^R_{hh}=\tilde{D}^R\tilde{\mathcal{S}}_\mathcal{D}$ & $\overline{t}^R_{hh}=\tilde{D}^R\tilde{\mathcal{S}}_\mathcal{D}$\\
  		\hline\hline
  	\end{tabular}
  \end{table}
\subsection{Electronic transport}
\label{App_trans}
Consider a junction between a superconductor (S) and a normal metal (N), see Fig. \ref{Fig_junc}.  
The superconductor is situated at $z<0$ and the normal metal at $z>0$. 
Assuming a point contact between the two (approximately) transversely invariant N and S regions, we reduce the problem to variations in only one spatial dimension, the longitudinal $z$-direction.
\subsubsection{Interface scattering matrix}
Now we will briefly describe the theoretical model of the NS interface. Let us imagine that some of the magnetic impurities, residing in the superconductor, are pinned to the NS surface. 
In this case tunneling through the interface would become, in general, spin-dependent \footnote{It is worth noting that we assume the interface and bulk impurity magnetic moments to be collinear in the ferromagnetically aligned case.}.
This can be simulated using the spin-active interface model \cite{Tokuyasu1988,Fogelstrom2000,Zhao2004}, with a normal-state electron scattering matrix (evaluated at the Fermi energy) of the form
\begin{equation}
S_e=
\begin{pmatrix}
\mathcal{S}_\mathcal{R} & \mathcal{S}_\mathcal{D}\\
\mathcal{S}_\mathcal{D} & -\mathcal{S}_\mathcal{R}
\end{pmatrix},
\label{SM_model}
\end{equation}
where
\begin{equation}
\begin{aligned}
\mathcal{S}_\mathcal{X}=&
\begin{pmatrix}
\sqrt{\mathcal{X}_\uparrow}e^{i \frac{\Theta}{2}} & 0\\
0 & \sqrt{\mathcal{X}_\downarrow}e^{-i \frac{\Theta}{2}}
\end{pmatrix},\;\;
\mathrm{\mathcal{X} = \{\mathcal{R},\mathcal{D}\}}.
\end{aligned}
\end{equation}
The scattering matrix for holes is related to the electron one through $S_h=\tilde{S}_e^{\dagger}$, with the ``tilde''-operation defined in Eq. \eqref{tilde_def}. Scattering probabilities $\mathcal{D}_{\sigma}$ and $\mathcal{R}_{\sigma}$ ($\sigma=\uparrow,\downarrow$) satisfy the usual conservation law, $\mathcal{D}_{\sigma}+\mathcal{R}_{\sigma}=1$.
In this model, besides unequal spin-resolved transmission probabilities, quasiparticles can acquire a spin-dependent phase shift $\Theta_{\uparrow,\downarrow}$. The latter property enters via the so-called spin-mixing angle $\Theta = \Theta_{\uparrow} - \Theta_{\downarrow}$. 

In order to completely determine our model we have to make assumptions about the dependence of tunneling probabilities on the quasiparticle momentum direction $\mathbf{p}_{F}$. We utilize two models, either an angle independent transmission function or an angle dependence derived from a $\delta$-function interface potential \cite{Blonder1982}
\begin{equation}
\mathcal{D}(\theta)=\left\{
\begin{array}{ll}
\mathcal{D}_0\:\forall\:\theta, & (\mbox{angle-independent}),\\
\displaystyle \frac{\mathcal{D}_0\cos^2\theta}{1-\mathcal{D}_0\sin^2\theta}, & (\delta-\mbox{function}).
\end{array}\right.
\label{delta_barrier}
\end{equation} 
Here, $\theta$ is the incidence angle, which is the angle between a quasiparticle's momentum and the normal to the NS surface, and $\mathcal{D}_0$ is the transmission probability at normal incidence.

The normal state scattering matrix completely determines transport properties of the NS interface in the superconducting state, if we use the quasiclassical boundary conditions \cite{Eschrig2000,Fogelstrom2000,Zhao2004,Eschrig2009}. They relate the quasiclassical Green's functions on the incoming quasiparticle trajectories to the outgoing ones.
We do not write them here for brevity, but rather suggest the interested reader to look into the original references.
	\begin{figure}[t]
		\centering
		\includegraphics[width=0.65\columnwidth,keepaspectratio]{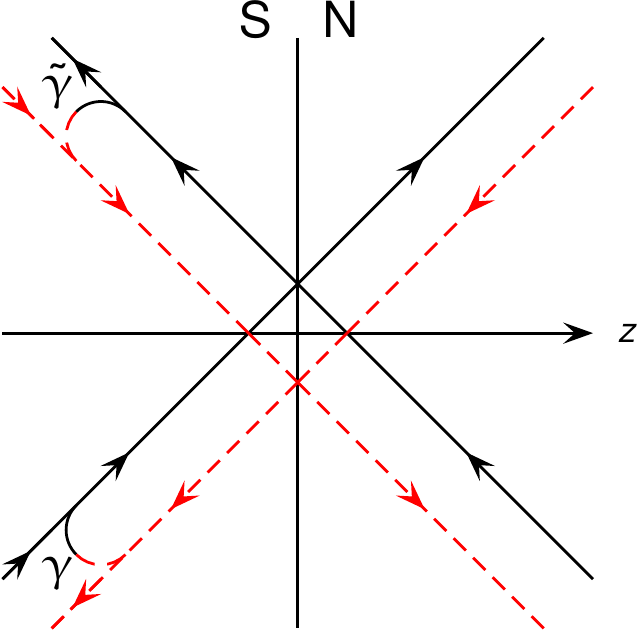}
		\caption{\label{Fig_AS}%
			Diagram illustrating all possible elementary processes occurring at an NS interface. 
			The full black (red dashed) lines denote electron-like (hole-like) quasiparticle trajectories. The arcs connecting full and dashed lines show the possibility of particle-hole branch conversion (Andreev reflection). The coherence functions $\gamma$ and $\tilde{\gamma}$ have a meaning of probability amplitude of $h \rightarrow e$ and $e \rightarrow h$ conversions, respectively \cite{Eschrig2009}. To construct arbitrary amplitude, one starts with an incoming (arrow pointing towards the interface) particle or hole line, traces all possible ways to arrive at the desired outgoing line, and takes a superposition of them.
		}%
	\end{figure}
	\begin{figure*}[!htb]
		\centering
		\includegraphics{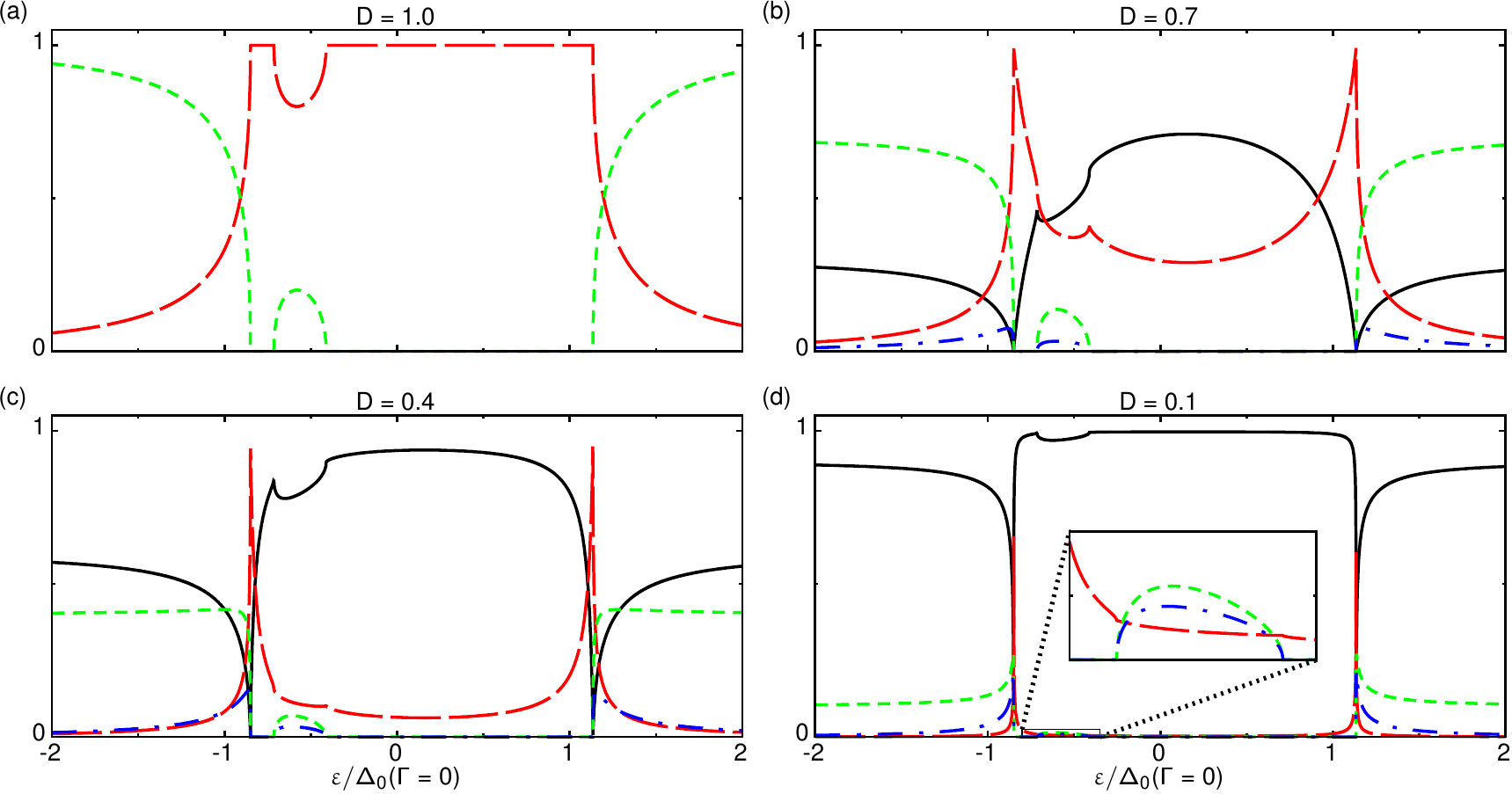}
		\caption{\label{Fig_SC}%
			Scattering amplitudes at the NS interface for spin-up quasiparticles, calculated for different transparencies, see Eq.~(\ref{SC_prob2}). Impurity spins are assumed ferromagnetically aligned with $\alpha>0$, and other parameters are taken as: $\Gamma/2 \pi k_B T_{c0}=0.01$, $u_{\mathrm{S}}=4$, and $T=0.01 T_{c0}$. Here, $T_{c0}$ is the clean-limit critical temperature of the superconductor (at $\Gamma = 0$). The black full line is the probability of normal reflection, $R_{ee,\uparrow}$, while the red long-dashed line corresponds to Andreev reflection, $R_{he,\uparrow}$. The green short-dashed and the blue dash-dotted lines are the normal transmission, $T_{ee,\uparrow}(1-|\tilde{\gamma}_\uparrow|^2)$, and transmission with branch conversion, $T_{he,\uparrow}(1-|\gamma_\uparrow|^2)$, respectively. The inset in (d) is a zoom on energies around the YSR impurity band. We also note the effective Zeeman shift [$\sim 0.04 \times 2 \pi k_B T_{c0}\approx 0.15 \Delta_0 (\Gamma=0)$ for current parameters] due to ferromagnetically ordered impurity spins, see Eq.~(\ref{SE_aligned}).
			The interface is assumed spin-inactive here ($\Theta=0$ and $\mathcal{D}_{\uparrow}=\mathcal{D}_{\downarrow}$) and the transmission probabilities independent of the quasiparticle incidence angle $\theta$, see Eq.~(\ref{delta_barrier}).
		}%
	\end{figure*}	
  \subsubsection{Elementary scattering processes}
  Transport across the NS junction can be described in terms of a few elementary processes taking place at the interface. 
  Each elementary process is described by the corresponding scattering amplitude. We summarize all possible amplitudes in Table \ref{scatter_amplitudes2}, and note that they are $2\times2$ matrices in spin space. The amplitudes have a clear physical meaning. For example, $r_{\alpha\beta}$ ($t_{\alpha\beta}$) is the probability amplitude of a $\beta$ excitation incident from N to be reflected (transmitted) as an $\alpha$ excitation. Here, $\alpha$ and $\beta$ can be either $e$ or $h$, referring to electron-like and hole-like quasiparticles, respectively. Another way to understand these amplitudes is by considering the diagram depicted in Fig. \ref{Fig_AS}. Each amplitude corresponds to an infinite series of Feynman paths a quasiparticle can take in order to get from the initial incident trajectory to the final outgoing one, with possibly undergoing a particle-hole conversion (Andreev reflection). 
  
  Similarly to the advanced and retarded propagators we had above, we can introduce advanced scattering amplitudes, related to the retarded ones via
	\begin{align}
		r^A_{\alpha\beta} = \left[r^R_{\alpha\beta}\right]^{\dagger}, t^A_{\alpha\beta} = \left[t^R_{\alpha\beta}\right]^{\dagger},
	\end{align}
  where Hermitian conjugation operates only in spin space. The amplitudes in Table \ref{scatter_amplitudes2} obey a number of relations, and, in particular, they satisfy the following relation \cite{Lofwander2003},
	\begin{align}
    R_{ee,\sigma}+R_{he,\sigma}&+T_{ee,\sigma}(1-|\tilde{\gamma}^R_{\sigma}|^2)\notag\\
    &+T_{he,\sigma}(1-|\gamma^R_{\sigma}|^2) = 1, \;\; \sigma = \left\{\uparrow,\downarrow\right\},
	\label{SC_prob2}
	\end{align}
  which is a manifestation of current conservation across the interface. Here we have defined the scattering probabilities, which are related to the corresponding amplitudes as
    \begin{equation}
    R_{\alpha\beta,\sigma}=|r^R_{\alpha\beta,\sigma}|^2, \;\; T_{\alpha\beta,\sigma}=|t^R_{\alpha\beta,\sigma}|^2.
    \label{SC_probs}
  \end{equation}
  On the lhs of Eq.~(\ref{SC_prob2}), the first term describes normal reflection of an incident (from N) electron-like quasiparticle, while the second one refers to reflection with a $e \rightarrow h$ branch conversion. The third and fourth terms are the corresponding transmission processes, which have additional prefactors though. The latter can be understood heuristically as the probabilities of staying an electron-like, $(1-|\tilde{\gamma}^R_{\sigma}|^2)$, or a hole-like, $(1-|\gamma^R_{\sigma}|^2)$, quasiparticle in S after being transmitted across the interface, without eventually getting branch-converted. All these terms can be easily identified with those obtained by Blonder, Tinkham and Klapwijk \cite{Blonder1982} using an alternative wavefunction matching approach.
  
  It is instructive to look at the energy dependence of the four scattering probabilities discussed above. In Fig.~\ref{Fig_SC} we plot them for spin-up quasiparticles in the case of ferromagnetically aligned impurity spins and various transparencies of the NS interface, which in this case is not spin-active. Corresponding plots for spin-down quasiparticles are obtained by mirroring each plot in Fig.~\ref{Fig_SC} with respect to $\varepsilon=0$, and are therefore not shown here. If impurity spins are oriented randomly, the extra sub-gap structure, related to the band of YSR states, would be present symmetrically at both positive and negative energies. For a completely transparent interface ($\mathcal{D} = 1$), see Fig.~\ref{Fig_SC}(a), at sub-gap energies the only allowed process is Andreev reflection, except for energies corresponding to the YSR band. In the latter case, there is also a small fraction of direct single-particle tunneling, however Andreev reflection dominates. For energies larger than the superconducting gap, the single-particle processes become dominant very quickly, as the energy is increased. When the interface transparency decreases, the two other processes come into play, namely normal reflection and transmission with a branch conversion ($e \rightarrow h$ or $h \rightarrow e$). They both require non-zero normal-state reflection, $\mathcal{R} > 0$, at the interface, see Table~\ref{scatter_amplitudes2} and Fig.~\ref{Fig_AS}. All probabilities, except for normal reflection $R_{ee,\uparrow}$, decrease when $\mathcal{D}$ decreases, but with a different rate. Indeed, by looking at the definition of scattering amplitudes in Table~\ref{scatter_amplitudes2}, one can see that Andreev reflection probability decreases faster than the others, $R_{he,\uparrow}\propto \mathcal{D}^2$. It means that in the tunneling limit, $\mathcal{D}\ll1$, the dominant transfer process is single-particle tunneling, see Fig.~\ref{Fig_SC}(d).

\subsubsection{Charge and spin currents}
	Let us now discuss the technical details of calculating spin and charge currents in our setup. The general expression for the electric current (injected in the $z$-direction) reads
	\begin{equation}
    I^c = \frac{eN_FA_c}{8\pi i}\int\limits_{-\infty}^{\infty}\mathrm{d}\varepsilon
    \int\frac{\mathrm{d}\Omega_{\mathbf{p}_F}}{4\pi}\mathrm{Tr}\left[v_{Fz}\hat{\tau}_3\hat{g}^K(\varepsilon,\mathbf{p}_F)\right],
		\label{c_current}
	\end{equation}
	where $v_{Fz} = v_F\cos\theta$ is the $z$-component of Fermi velocity (incidence angle $\theta$ is the polar angle of a coordinate system with the $z$-axis normal to the interface), $e$ is the electron charge, and $A_c$ is the contact area. Here the Keldysh component of the Green's function, ${g}^K(\varepsilon,\mathbf{p}_F)$, is computed at the interface, $z = 0$. Trace is taken over both Nambu and spin spaces. As discussed above, in our problem we can introduce spin-resolved quantities, which allow us to define both charge and spin currents in a usual way,
	\begin{align}
	I^c = I_{\uparrow} + I_{\downarrow},\;\;I^s = I_{\uparrow} - I_{\downarrow}.
	\end{align}
	For the case of randomly oriented impurity spins, if the interface scattering is spin-independent ($\Theta=0$ and $\mathcal{D}_{\uparrow}=\mathcal{D}_{\downarrow}$), we have $I_{\uparrow} = I_{\downarrow}$, and consequently $I^s = 0$.

  If we assume the normal side of the interface to be disorder-free, all incoming coherence functions from N vanish (because there is no order parameter in bulk N). Then we can write down the incoming Keldysh Green's function computed at the interface in N, $z = 0^{+}$ (see Fig.~\ref{Fig_AS}), as
	\begin{equation}
		\hat{g}^K_{N,in}=-2\pi i
		\begin{pmatrix}
			x_N & r^A_{he}x_N\\
			-r^R_{he}x_N & \tilde{X}_N-r^R_{he}r^A_{he}x_N
		\end{pmatrix},
		\label{g_K_in}
	\end{equation}
	while the outgoing one has the form
	\begin{equation}
		\hat{g}^K_{N,out}=-2\pi i
		\begin{pmatrix}
			X_N-r^R_{eh}r^A_{eh}\tilde{x}_N & -r^R_{eh}\tilde{x}_N\\
			r^A_{eh}\tilde{x}_N & \tilde{x}_N
		\end{pmatrix}.
		\label{g_K_out}
	\end{equation}
	Here, $x_N$ and $\tilde{x}_N$ are the distribution functions of the incoming electron-like and hole-like quasiparticles, while 
	$X_N$ and $\tilde{X}_N$ are their outgoing counterparts. The latter can be written in terms of the former as \cite{Eschrig2000}
	\begin{equation}
	\begin{aligned}
	&X_N=r^R_{ee}r^A_{ee}x_N+\overline{t}^R_{ee}x_S\overline{t}^A_{ee}-\overline{t}^R_{eh}\tilde{x}_S\overline{t}^A_{eh},\\
	&\tilde{X}_N=r^R_{hh}r^A_{hh}\tilde{x}_N+\overline{t}^R_{hh}\tilde{x}_S\overline{t}^A_{hh}-\overline{t}^R_{he}x_S\overline{t}^A_{he},
	\end{aligned}
	\label{X_out}
	\end{equation}
	where $x_S$ and $\tilde{x}_S$ are the distribution functions for quasiparticles incident from S. We assume that the incoming distribution functions take their bulk values, so that 
	\begin{equation}
		\begin{aligned}
			&x_N=\tanh[(\varepsilon-eV)/2k_BT],\\
			&x_S=(1-\gamma^R\tilde{\gamma}^A)\tanh[\varepsilon/2k_BT],
		\end{aligned}
		\label{X_in}
	\end{equation}
	where $V$ is the bias, $k_B$ is the Boltzmann constant, and the ``tilded'' counterparts are found via Eq.~(\ref{tilde_def}). Plugging equations \eqref{g_K_in}-\eqref{X_out} into Eq. \eqref{c_current} we obtain
	\begin{equation}
    I_{\uparrow,\downarrow} = -\frac{eN_FA_c}{4}\int\limits_{-\infty}^{\infty}\mathrm{d}\varepsilon\int\frac{\mathrm{d}\Omega_{\mathbf{p}_F}}{4\pi}\mathrm{Tr}\Big[v_{Fz}j_{\uparrow,\downarrow}\Big],
	\label{current_final}
	\end{equation}
  where we have defined
	\begin{align}
		j_{\sigma} &= (1-R_{ee,\sigma}+R_{he,\underline{\sigma}})x_N + (\overline{T}_{he,\underline{\sigma}}x_{S,\underline{\sigma}} -
		\overline{T}_{ee,\sigma}x_{S,\sigma})\notag\\
		&+ (1-R_{hh,\underline{\sigma}}+R_{eh,\sigma})\tilde{x}_N + (\overline{T}_{eh,\sigma}\tilde{x}_{S,\sigma} - 
		\overline{T}_{hh,\underline{\sigma}}\tilde{x}_{S,\underline{\sigma}}),\notag\\
		&\hspace{0.3\columnwidth}\sigma = \left\{\uparrow,\downarrow\right\},\;\;\underline{\uparrow} = \downarrow.
	\label{j_sigma}
	\end{align}
  The form of spin-resolved currents might look confusing at first glance, seemingly mixing the spin channels. However, one has to remember that working in a $4\times 4$ Nambu-spin space introduces some redundancy in the formalism by dealing with both particles and holes of two spin flavors. A hole quasiparticle carries positive charge and propagates in the opposite direction to an electron quasiparticle. Therefore, one can intuitively understand the hole-related part of the spin-resolved current in Eq.~(\ref{j_sigma}). Another way to look at Eqs.~(\ref{current_final})-(\ref{j_sigma}) is to get rid of the hole-related terms by working in the excitation picture. In order to do it, one has to perform a transformation $\varepsilon\rightarrow -\varepsilon$ for ``tilded" terms in Eq.~(\ref{j_sigma}), in which case they simply double the particle terms.
  
  Let us demonstrate how it works in a simple case of a $s$-wave spin-singlet superconductor placed in a Zeeman exchange field $H$. In this case, the density of quasiparticle states in the bulk has the form,
  \begin{equation}
  \begin{gathered}
	  N_{\uparrow,\downarrow}(\varepsilon)=\mathrm{Im}\left[\frac{\epsilon\mp\varepsilon_Z}{\sqrt{\Delta_0^2-(\epsilon\mp\varepsilon_Z)^2}}\right],\\
	  \varepsilon_Z = \frac{1}{2}g\mu_B H,\;\;\epsilon = \varepsilon + i\eta,\;\;\eta\rightarrow 0^{+},
  \end{gathered}
  \end{equation}
  where $\mu_B$ is the Bohr magneton and $g\approx 2$ is the $g$-factor. Then, if we assume that $\mathcal{D}\ll 1$, corresponding to
  the tunneling limit, as studied by Merservey and Tedrow\cite{Meservey1970,Meservey1994},
  we can write down, for example, $j_{\uparrow}$ to linear order in $\mathcal{D}$ as follows,
  \begin{align}
  j_{\uparrow} = &-2\mathcal{D}N_\uparrow(\varepsilon)\left[f_{F}\left(\varepsilon-eV\right)-f_{F}\left(\varepsilon\right)\right]\notag\\
  &+2\mathcal{D}N_\downarrow(\varepsilon)\left[f_{F}\left(\varepsilon+eV\right)-f_{F}\left(\varepsilon\right)\right],
  \label{j_up_Zeeman}
  \end{align}
  where $f_{F}(\varepsilon)=\left[1+\text{exp}\left(\varepsilon/k_B T\right)\right]^{-1}$ is the Fermi distribution function. The second term on the rhs of Eq.~(\ref{j_up_Zeeman}) corresponds to the hole (``tilded") terms in Eq.~(\ref{j_sigma}). If we transform $\varepsilon\rightarrow -\varepsilon$ in this term, and use the fact that $N_{\downarrow}(-\varepsilon) = N_{\uparrow}(\epsilon)$ and $f_{F}(-\varepsilon) = 1-f_{F}(\varepsilon)$, we simply get the first term on the rhs of Eq.~(\ref{j_up_Zeeman}). This simple example demonstrates that the concept of a hole is convenient for doing calculations, but it does not alter the usual logic of charge and spin currents known for normal (non-superconducting) systems. 
 
  Finally, in order to quantify the degree of spin polarization of the electric current, we define the following quantity, 
  \begin{equation}
    P=\frac{G^s}{G^c},
    \label{P_def}
  \end{equation}
  where $G^{c(s)}=dI^{c(s)}/dV$ is the differential charge (spin) conductance. When $P = 1$ ($P = -1$) the current is carried only by spin-up (spin-down) quasiparticles.
\subsubsection{Current noise and differential Fano factor}
In this section we briefly describe how to calculate the current noise and define the differential Fano factor. The noise is expressed via a current-current correlation function, and our derivation closely follows Ref.~[\onlinecite{Lofwander2003}]. The final expression consists of two terms: one coming from products of Keldysh Green's functions, and another one originating from products of retarded-advanced Green's functions (the cross-terms vanish). It can be written as
\begin{equation}
\mathcal{S} = \frac{e^2N_FA_c}{4}\int\mathrm{d}\epsilon\int\frac{\mathrm{d}\Omega_{\mathbf{p}_F}}{4\pi} v_{Fz}\mathrm{tr}_\sigma\left[\mathcal{S}^K-\mathcal{S}^{R-A}\right],
\label{noise_eq}
\end{equation}
where trace is taken over spin space. Expression for $\mathcal{S}^K$ and $\mathcal{S}^{R-A}$ can be found in Appendix~\ref{App1}. 

When we know how to compute both the electric current and noise, we can define the differential Fano factor,
\begin{equation}
	F=\frac{1}{2eG^c}\frac{d\mathcal{S}}{dV}.\label{Fano_eqn}
\end{equation}
We note that $F$ can be measured directly, see for example Ref.~[\onlinecite{Schoelkopf1997}], or obtained from a $\mathcal{S}(V)$ measurement.
  \section{Results}
  \label{Sec_Results}
  In this section we present the results of our numerical calculations for the three different cases mentioned above. 
  We start by discussing transport across an NS junction in the tunneling regime. 
  Next, we consider how transport characteristics of the junction evolve as transparency is increased. 
  Finally, allowing the NS interface to be spin-active, we investigate the role of inverse proximity effect on transport.

  \subsection{Tunneling regime}
	In the tunneling regime, $\mathcal{D}\ll1$, currents across the NS interface due to the applied bias are small. This allows us to neglect changes in the superconducting order parameter and consider it spatially constant. Then, the incoming coherence functions $(\gamma^X,\tilde\gamma^X,\,X=R,A)$ are given by their bulk values, and we can easily compute transport characteristics of the system, without a need to resort to full self-consistent calculations of spatially varying self-energies. Note that the bulk self-energies are still computed self-consistently as in Ref.~\onlinecite{Persson2015}.
\begin{figure}[t]
	\centering
	\includegraphics[width=0.82\columnwidth]{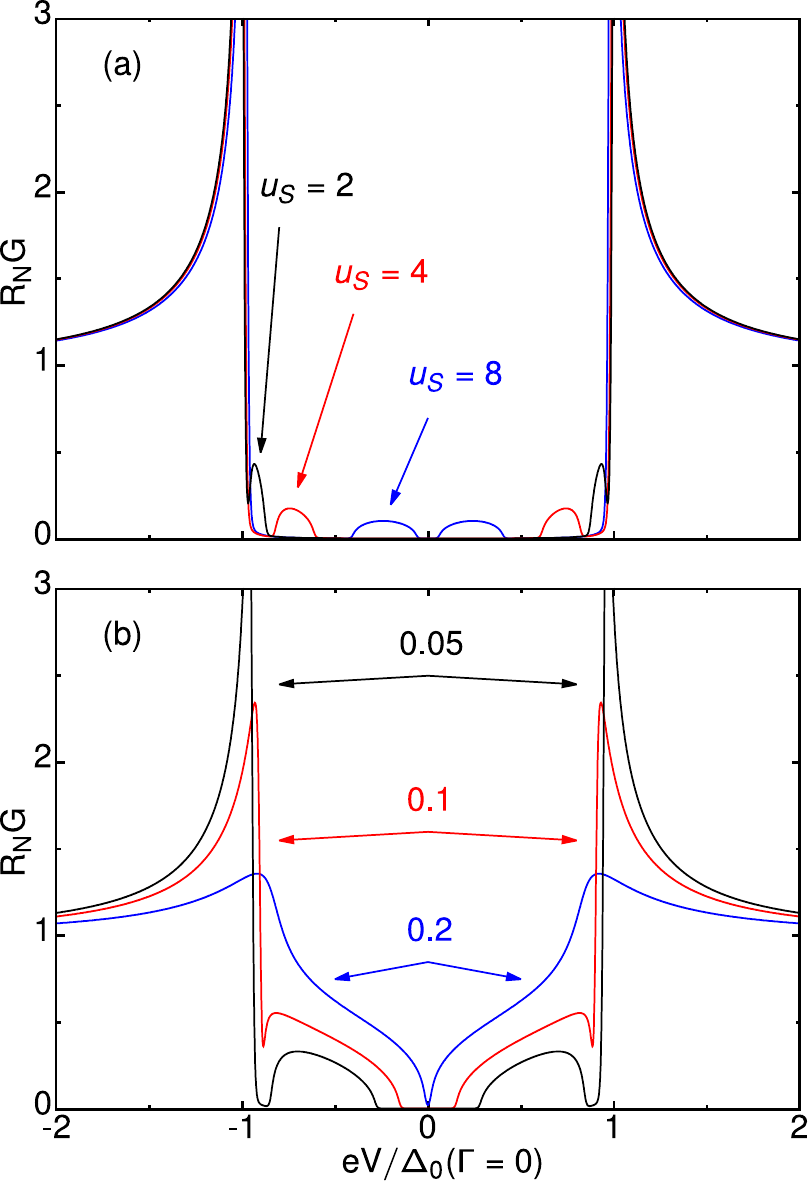}
	\caption{Differential conductance for the case of randomly oriented impurity spins, given in units of normal-state conductance $G_N$, where $G_N = R_N^{-1} = 2e^2N_FA_c\mathcal{D}\langle v_{Fz}\rangle_{\mathbf{p}_F\cdot\mathbf{e}_z < 0}$. Here, $\langle v_{Fz}\rangle_{\mathbf{p}_F\cdot\mathbf{e}_z < 0}$ is the average velocity of quasiparticles with momenta pointing towards S, see Fig.~\ref{Fig_AS}. Panel (a) shows the effect of increasing $u_\mathrm{S}=2, 4, 8$ with 
		$\Gamma/2\pi k_B T_{c0}=0.01$. Panel (b) demonstrates the effect of increasing $\Gamma/2\pi k_B T_{c0}=0.05,0.1,0.2$ with $u_\mathrm{S}=5$. The interface parameters are: $\Theta = 0$ and $\mathcal{D}_{\uparrow}=\mathcal{D}_{\downarrow}=0.01$. Transmission probabilities are assumed independent of the incidence angle $\theta$, see Eq.~(\ref{delta_barrier}). The temperature is $T=0.01 T_{c0}$, where $T_{c0}$ is the clean-limit critical temperature of the superconductor (at $\Gamma = 0$).}
	\label{conductance_mso0_tunnel}
\end{figure}
\begin{figure}[t]
	\centering
	\includegraphics[width=0.82\columnwidth]{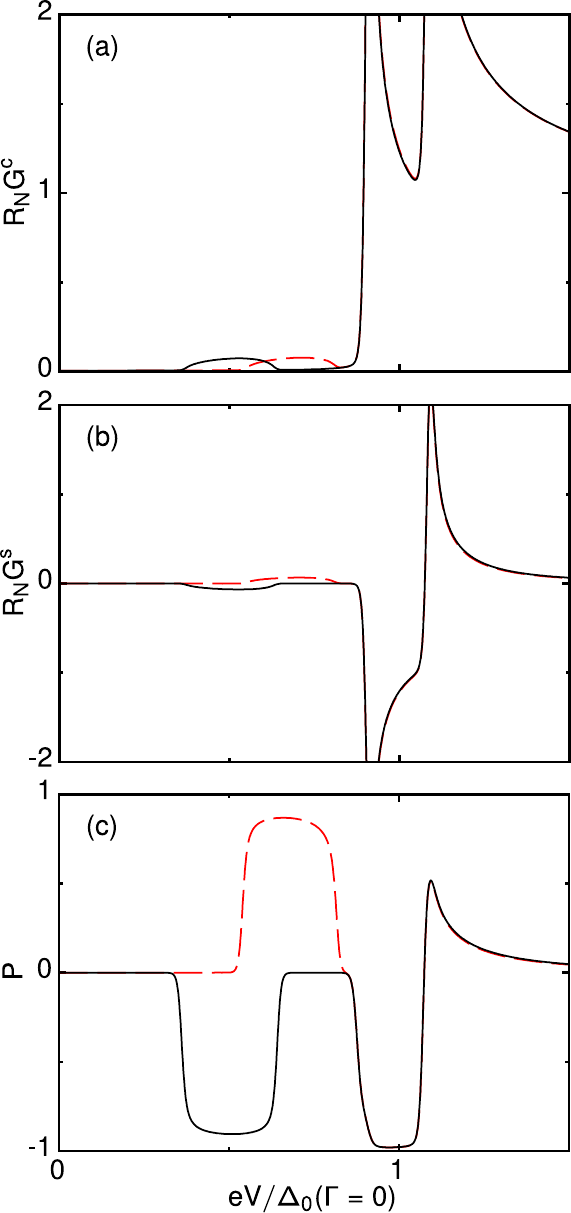}		
	\caption{Charge (a) and spin (b) differential conductance as a function of bias. Panel (c) represents their ratio $P$ [see Eq.~(\ref{P_def})], providing information on the spin-polarization of current. Impurity spins are assumed ferromagnetically aligned. Black full lines correspond to $\alpha>0$ and red dashed ones to $\alpha<0$. The system parameters are  $u_\mathrm{S}=5$ and $\Gamma/2\pi k_B T_{c0}=0.005$. The interface parameters and temperature are the same as in Fig.~\ref{conductance_mso0_tunnel}.}
	\label{conductance_mso1_tunnel}
\end{figure}
\begin{figure}[t]
	\centering
	\includegraphics[width=0.82\columnwidth]{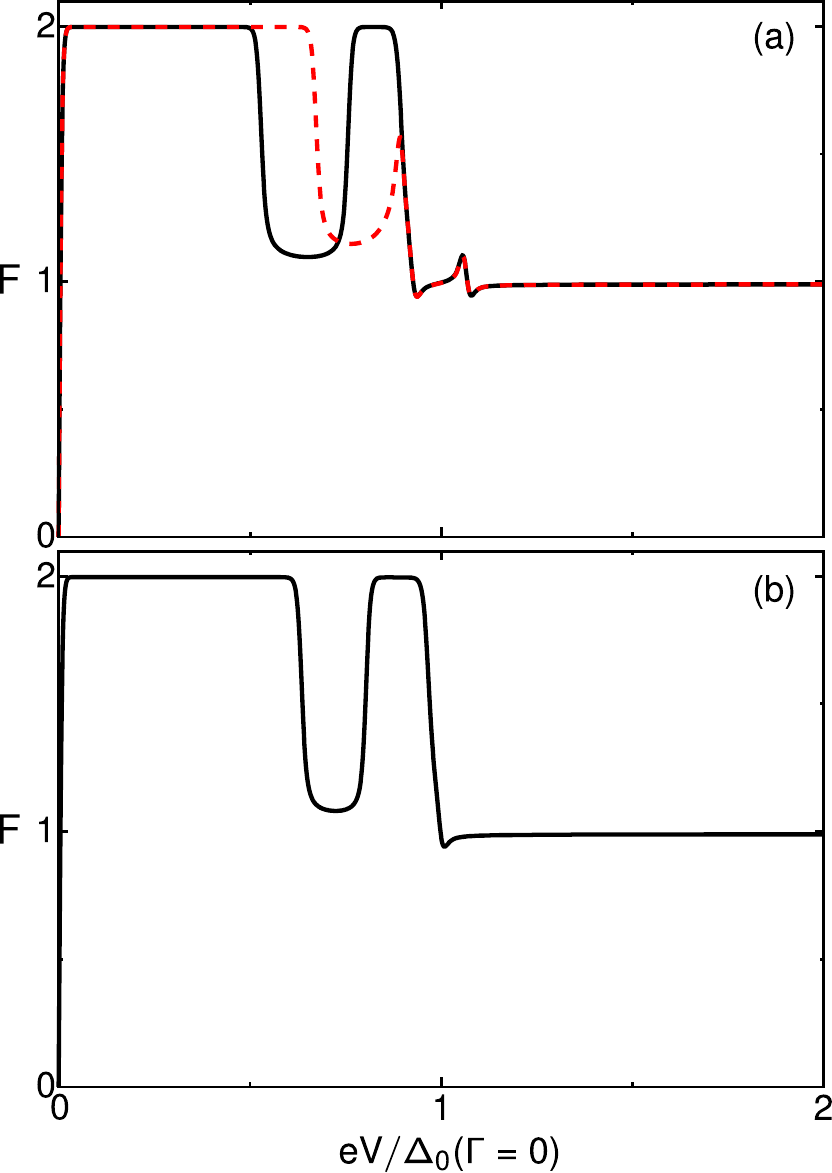}
	\caption{Differential Fano factor for the case of ferromagnetically ordered (a) and randomly oriented (b) impurity spins. Black full lines correspond to $\alpha>0$, while red dashed ones to $\alpha<0$. The impurity strength is $u_\mathrm{S}=4$ and the other parameters are the same as in Fig.~\ref{conductance_mso1_tunnel}.}
	\label{fano_tunnel}
\end{figure}
\begin{figure*}[t]
	\centering
	\includegraphics[width=\textwidth]{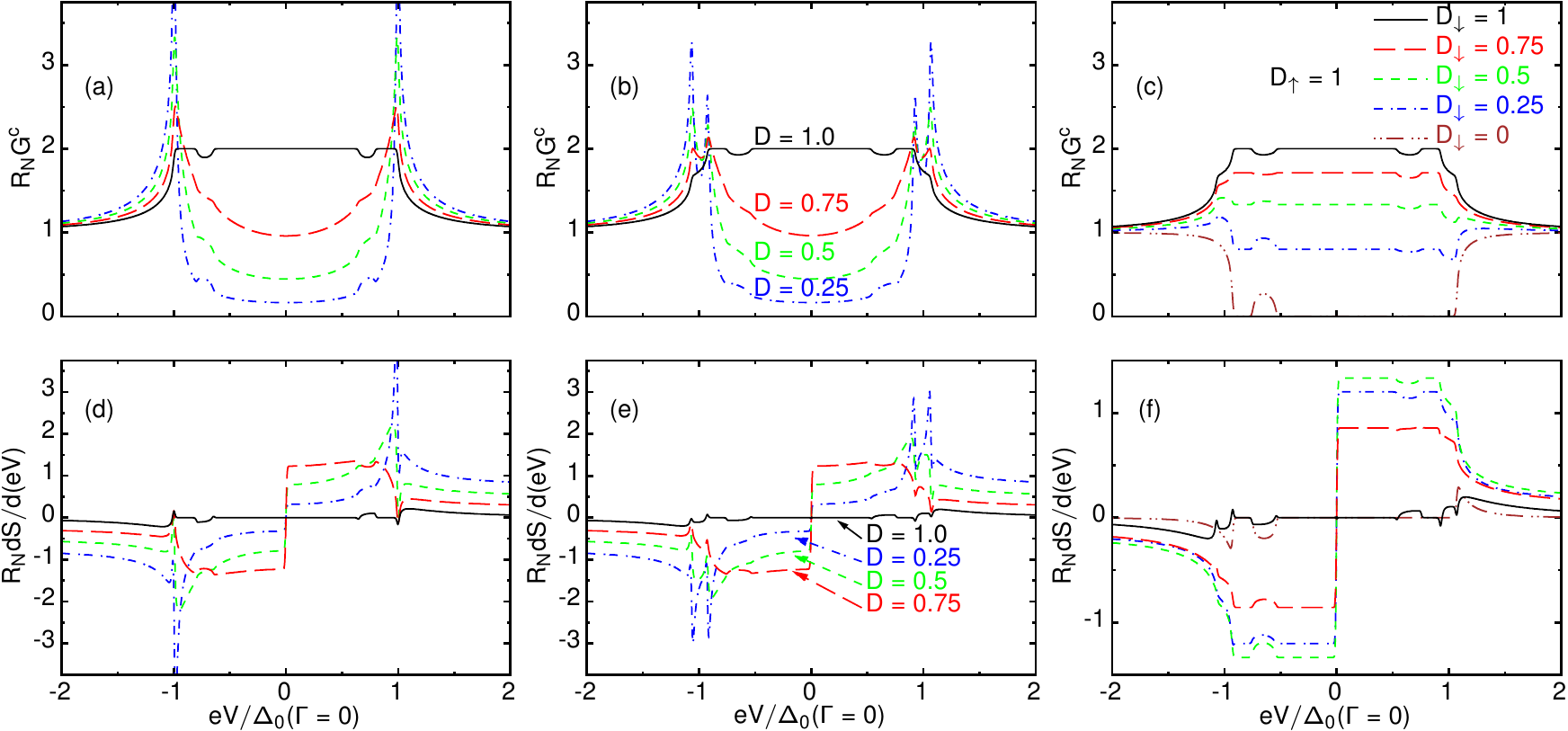}
	\caption{\label{conductance_noise}%
		Differential charge conductance and noise for the case of magnetic impurity spins being randomly oriented, (a) and (d), ferromagnetically aligned with $\alpha>0$, (b) and (e), and ferromagnetically aligned with $\alpha>0$ and spin-dependent transparency, (c) and (f). The system parameters are: $\Gamma/2 \pi k_B T_{c0}=0.005$, $u_{\mathrm{S}}=4$, $\Theta = 0$, and $T=0.01 T_{c0}$. Transmission probabilities are assumed independent of the incidence angle $\theta$, see Eq.~(\ref{delta_barrier}).}%
\end{figure*}	
\subsubsection{Conductance}
	We begin our discussion by considering differential conductance for the case of randomly oriented impurity magnetic moments.
	Figure~\ref{conductance_mso0_tunnel}(a) shows how the tunneling conductance is affected by the increase of impurity potential strength, $u_\mathrm{S}$, while Fig.~\ref{conductance_mso0_tunnel}(b) demonstrates its evolution with impurity density, $\Gamma$.
	The potential amplitude sets the position of YSR bands inside the gap, while the impurity density sets their width. It is well-known that the tunneling conductance is intimately related to the local density of states in the system, which allows for the YSR band spectroscopy \cite{Bauriedl1981,Ji2008,Ruby2015}, see Ref.~[\onlinecite{Persson2015}]. Increasing the density of impurities, brings more YSR states below the gap, which eventually cover the whole sub-gap region. This demonstrates the pair-breaking effect of magnetic impurities and the onset of so-called gapless superconductivity, as was shown by Abrikosov and Gor'kov\cite{Abrikosov1960} in the Born limit.

  Let us consider now the case of ferromagnetically ordered magnetic impurities. Then, the YSR bands become spin-polarized, and the basic response to varying impurity potential strength or density is the same as discussed above. There is an extra ingredient though, which is the effective Zeeman shift of the spin-resolved differential conductances, induced by the collective magnetic field of the impurities, see Eq.~(\ref{SE_aligned}) and Ref.~[\onlinecite{Persson2015}]. We focus instead on the fact that one can observe spin currents in this case. In Figs.~\ref{conductance_mso1_tunnel}(a),(b) we show charge and spin differential conductances as a function of bias, respectively. In order to quantify the spin-polarization of current, we combine these two quantities by plotting their ratio $P$ instead, see Eq.~(\ref{P_def}) and Fig.~\ref{conductance_mso1_tunnel}(c). One can observe that, for a given choice of system parameters, the current is nearly completely carried by single-spin quasiparticles ($P\sim90\%$) for the bias window probing the YSR band. Polarization of the YSR band increases upon decreasing transparency ($P>99\%$ for $\mathcal{D}\lesssim 10^{-3}$). Another important feature is that the relative shift of spin-resolved conductances, induced by the collective magnetic field of the impurities, enables a very high degree of current spin-polarization for the bias window close to the Zeeman-split gap edges. In particular, for the case of ferromagnetic exchange between the impurity spins and the itinerant electrons ($\alpha<0$), there is a possibility to choose the electric current spin-polarization by simply tuning the bias around $eV/\Delta_0(\Gamma = 0)\simeq 1$, see red dashed line in Fig.~\ref{conductance_mso1_tunnel}(c). This feature makes such systems potentially suitable for on-demand production of quasiparticles with specific spin projection.

  \subsubsection{Differential Fano factor}
  So far, we have discussed only the spin polarization of electric current. In the tunneling regime, one can get an insight into the nature of elementary charge carriers and their statistics by looking at current fluctuations \cite{Blanter2000}. Namely, let us consider the differential Fano factor, as a measure of the carrier effective charge (in units of $e$). 
  In the normal (non-superconducting) state we have $F = 1$, indicating that the electric current is transfered by quasiparticles with the effective charge equal to $e$, see Ref.~[\onlinecite{Blanter2000}]. In superconductors, besides the single-particle tunneling, we also have Andreev reflection processes \cite{Andreev1964}, which imply a transfer of charge equal $2e$. So, the value of $F$ helps to decipher the relative role of single-particle and Andreev refection processes in the tunneling current.
	
  In Fig.~\ref{fano_tunnel} we plot the differential Fano factor for our setup. We observe that for all voltages above the gap $F = 1$, indicating the dominant role of single-particle tunneling. At voltages below the gap we have $F = 2$ for all energies except for those corresponding to the YSR impurity band, where we recover $F \approx 1$ again. This demonstrates, as expected, that Andreev reflection is the dominant mechanism of sub-gap electric transport (Cooper pair tunneling), if there are no single-particle states in this bias window. However, a sub-gap metallic impurity band (YSR band in our case), if present, predominantly supports single-particle tunneling.

\subsection{Metallic contact: high transparency}
In this section we investigate what happens to transport characteristics of the NS junction if we go beyond the tunneling limit. It is important to remind that we in this section assume a point contact between the superconductor and the normal-metal probe. It means that the contact area is much smaller than the superconducting coherence length, $A_c\ll\xi_0^2$. This circumstance allows us to assume the order parameter to be approximately spatially independent \cite{Fogelstrom2010,Fogelstrom2014}. 
  
  \subsubsection{Conductance and differential noise}
  If we look at the (charge) conductance and differential noise, see Fig.~\ref{conductance_noise}, we observe that there is no big difference between the cases of randomly oriented and ferromagnetically aligned impurity spins. The biggest noticeable difference is the Zeeman splitting in the latter case. For a completely transparent interface, $\mathcal{D} = 1$, the sub-gap differential conductance is equal to twice the normal-state value, except for the bias window which covers the YSR impurity bands, see Figs.~\ref{Fig_SC}(a)-(b). In the latter case, the probability of Andreev reflection slightly reduces, but it still remains the dominant mechanism of sub-gap transport, compared to single-particle tunneling. Decreasing the interface transparency, one observes a decrease of the sub-gap conductance, eventually recovering the tunneling regime discussed in the previous section. However, if we allow the transmission coefficient across the junction to be spin-dependent (spin-filtering effect), $\mathcal{D}_{\uparrow}\neq\mathcal{D}_{\downarrow}$, the conductance spectra look different, see Fig.~\ref{conductance_noise}(c). The change is caused by the fact that Andreev reflection in a spin-singlet superconductor requires quasiparticles of both spin flavors to have non-zero tunneling probability. In the extreme case, when one of the probablities vanishes, this process is forbidden.

  Let us now discuss the current noise, see Figs.~\ref{conductance_noise}(d)-(f). There are several sources of noise in a NS junction. Apart from the usual thermal Nyquist-Johnson noise (can be ignored at low temperatures), there is the shot noise, caused by fractional probabilities of single-particle tunneling and two-particle Andreev reflection. The latter feature makes the noise reach its maximal value at the interface transparency $\mathcal{D}_{\mathrm{max}}\simeq 0.75$ ($\mathcal{D}_{\mathrm{max}}\simeq 0.25$ above the gap), different from the normal-state value $\mathcal{D}_{\mathrm{max},N} = 0.5$ \cite{Blanter2000}, see Figs.~\ref{conductance_noise}(d)-(e). As we tune the spin-resolved transmission coefficients, weakening of Andreev reflection makes the differential noise acquire its maximum at transparencies approaching the normal-state value, see Fig.~\ref{conductance_noise}(f).

  \begin{figure}[t]
  \centering
    \includegraphics[width=0.95\columnwidth]{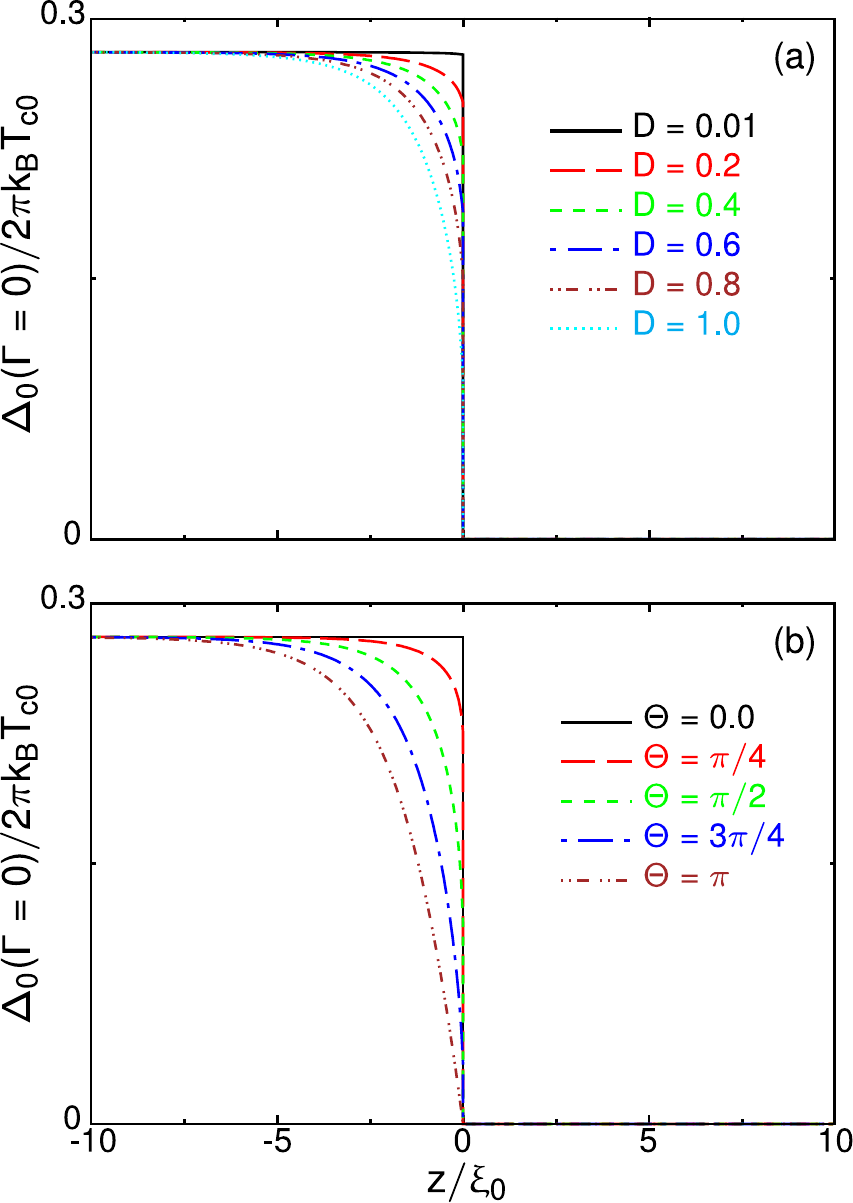}
    \caption{Spatial profile of the order parameter for a clean NS junction (no magnetic impurities) with a spin-active interface, see Fig.~\ref{Fig_AS}. (a) Effect of the interface transparency, $\mathcal{D}_{0\uparrow,\downarrow} = \mathcal{D}_0$ [see Eq.~(\ref{delta_barrier})] and $\Theta = 0$. (b) Effect of the spin-mixing angle, $\Theta\neq 0$ and $\mathcal{D}_{0\uparrow,\downarrow} = 0$. In both plots the temperature is $T=0.01T_{c0}$.}
	\label{inv_prox}
	\end{figure}
	
\subsection{Inverse proximity effect}
In the two previous sections we ignored the inverse proximity effect, i.e. a reduction of the superconducting order parameter in the vicinity of the NS interface.
In this section we discuss the impact of this effect on transport characteristics of the junction. In contrast to the previous sections, here we also consider an interplay between spin-activity of the interface, see Eqs.~(\ref{SM_model})-(\ref{delta_barrier}), and YSR impurity bands. Spin-active interfaces are known to host surface Andreev bound states (ABS) if $\Theta\neq 0$ \cite{Fogelstrom2000,Zhao2004,Hubler_PRL2012ABS}. These states appear below the gap at energies $\varepsilon^{\mathrm{ABS}}_{\uparrow,\downarrow}\approx\pm \Delta \cos(\Theta/2)$, where upper (lower) sign corresponds to spin-up (spin-down) quasiparticles.

In Fig.~\ref{inv_prox}(a) we plot the spatial profile of the order parameter for different values of the interface transparency. As expected, the order parameter reduction is higher for larger transparencies. If, on the other hand, the interface is completely insulating but $\Theta\neq 0$, one can observe the order parameter weakening due to formation of the ABS, see Fig.~\ref{inv_prox}(b).
\begin{figure*}[t]
\centering
\includegraphics[width=0.95\textwidth]{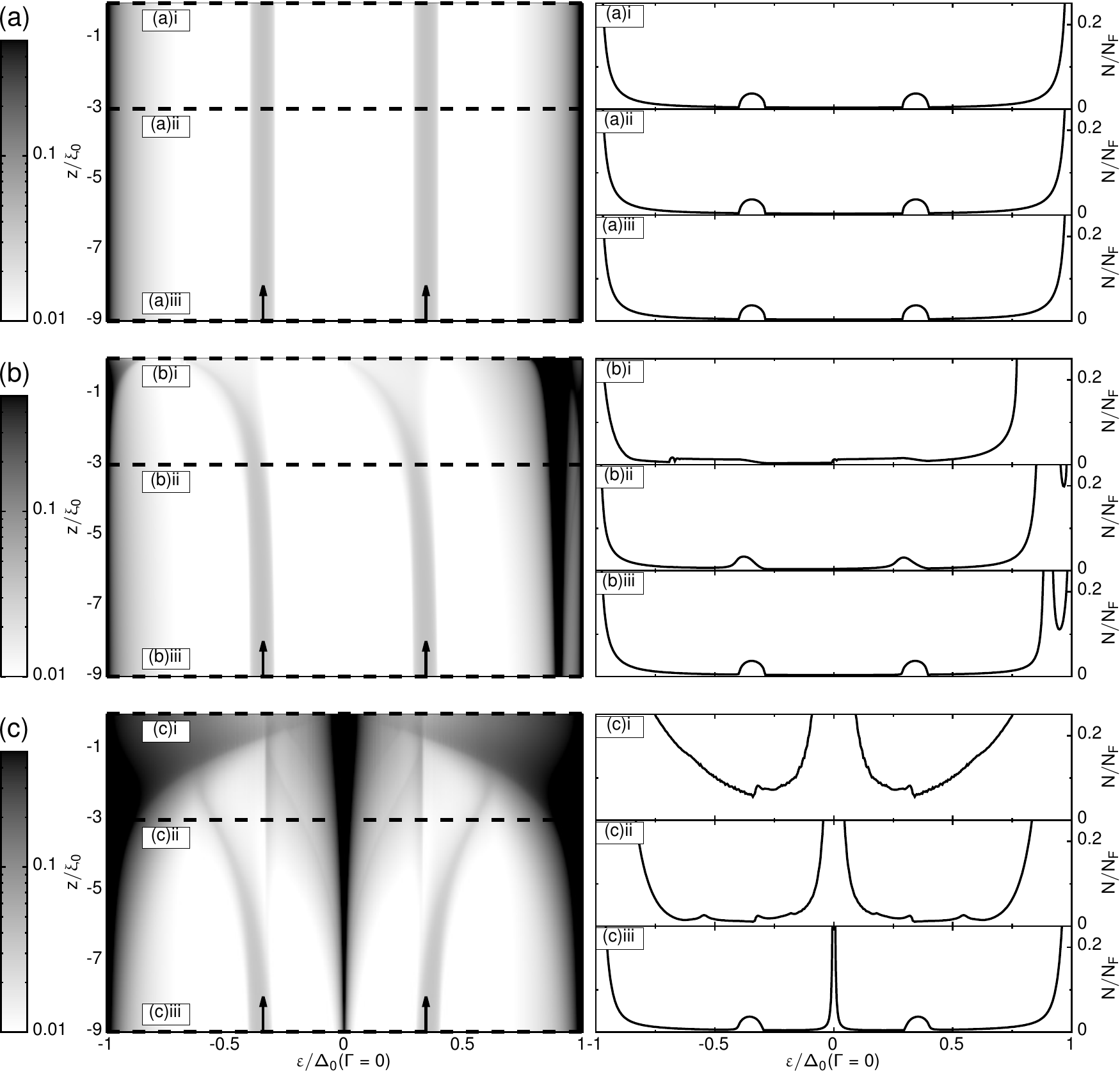}
\caption{
Local density of states in the superconductor for spin-up quasiparticles, assuming that magnetic impurities have randomly oriented spins. Left column shows heat maps of the density of states as a function of energy and distance from the interface, see Fig.~\ref{Fig_AS}. From top to bottom the spin-mixing angle is $\Theta=0,\pi/4,\pi$. Arrows indicate positions of the YSR bands in the bulk. Right column shows cuts of the heat maps at given values of distance from the interface, indicated by the dashed lines. For all plots we take $\Gamma/2 \pi k_B T_{c0}=0.001$, $u_{\mathrm{S}}=7$, $T=0.01 T_{c0}$, and $\mathcal{D}_{0\uparrow,\downarrow}=10^{-4}$ [see Eq.~(\ref{delta_barrier})].
}
\label{Dos_v_x}
\end{figure*}
	
Let us now come back to the case of an NS junction with magnetic impurities and illustrate the interplay of inverse proximity effect and the YSR impurity bands. In Fig.~\ref{Dos_v_x} we plot the local density of states in the superconductor as a function of energy and distance from the interface, see also Fig.~\ref{Fig_AS}. As can be seen from Fig.~\ref{Dos_v_x}(a), when $\Theta = 0$ there are no ABS at the NS surface and the YSR bands are unaffected by the inverse proximity effect. On the other hand, when $\Theta \neq 0$ the interface-induced ABS tend to repel and broaden (smear out) the YSR impurity bands. Indeed, in Fig.~\ref{Dos_v_x}(b) one can see that the ABS appears close to the gap edge and it repels the YSR bands away (towards negative energies) by smearing them out at the same time. The local density of states approaches its bulk form as we move away from the interface. Finally, when the ABS appears in between of the two YSR bands, the latter get repelled in opposite directions, see Fig.~\ref{Dos_v_x}(c).

So far we have discussed the impact of ABS on the YSR impurity bands. However, the shape of ABS gets modified as well by the presence of both impurities and non-zero interface transparency. The broadening of low-energy ABS due to impurities is known to range from $\propto\sqrt{\Delta\Gamma}$ in the Born limit, to $\propto\sqrt{\Delta\Gamma}{\mathrm{e}}^{-\Gamma/\Delta_0}$ in the unitary limit \cite{Fogelstrom2000,Kalenkov2004}. Finite interface transparency also contributes to the ABS broadening as $\propto \mathcal{D}\Delta$ \cite{Zhao2004}. 

We have to note that the results plotted in Fig.~\ref{Dos_v_x} were obtained for spin-up quasiparticles, assuming that the magnetic impurities have randomly oriented spins. The corresponding plots for spin-down case look exactly the same, but mirrored with respect to $\varepsilon=0$. For the case of spin-polarized impurities no new features appear. The YSR bands get spin-polarized and shifted by the collective Zeeman field of the impurities.
  
\begin{figure*}[t]
\centering
\includegraphics[width=\textwidth]{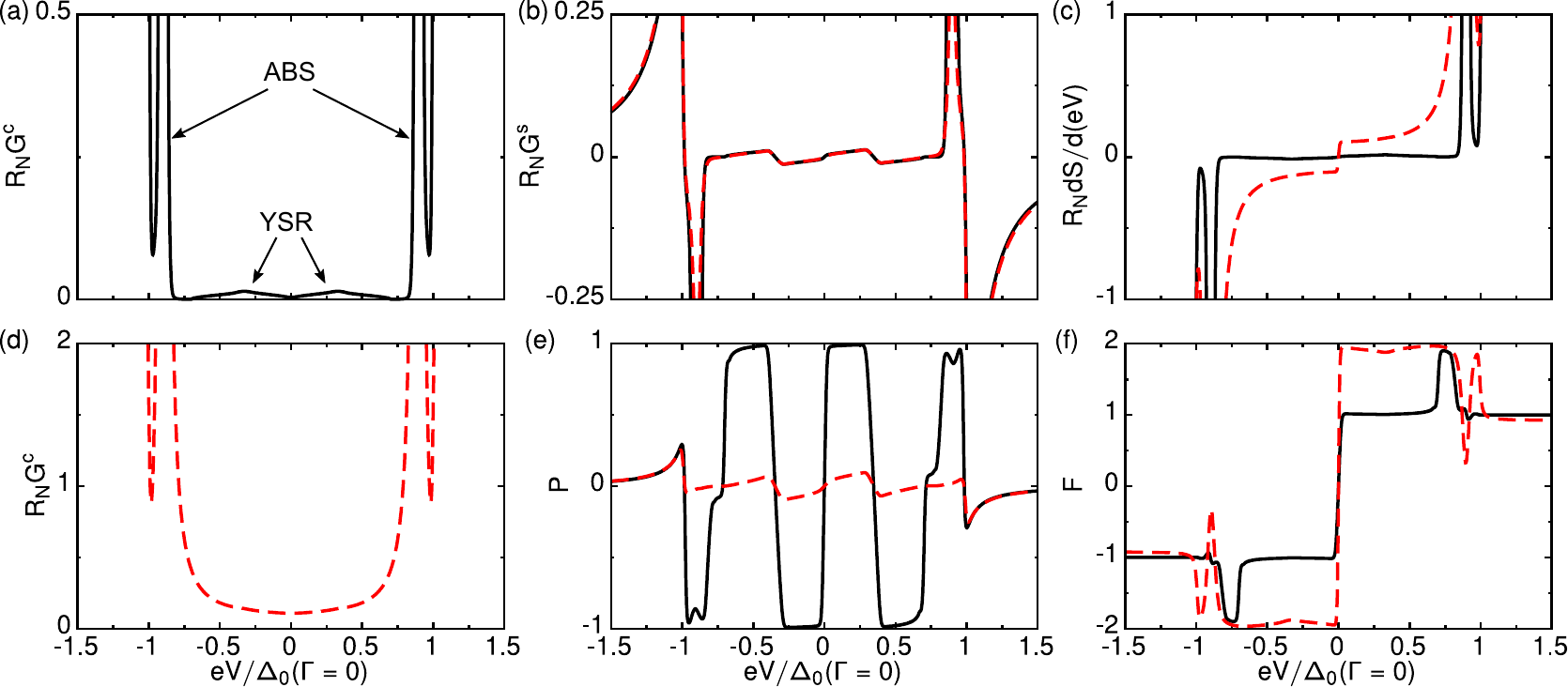}
\caption{Interplay of the ABS and YSR impurity bands. Impurities are assumed to have randomly oriented magnetic moments. The system parameters are: $\Gamma/2 \pi k_B T_{c0}=0.001$, $u_{\mathrm{S}}=7$, $T=0.01 T_{c0}$, and the spin-mixing angle is $\Theta=\pi/4$. The panels read: charge conductance, (a) and (d), spin conductance, (b), differential noise, (c), polarization, (e), and differential Fano factor, (f). Transparency of the interface is $\mathcal{D}_{0\uparrow,\downarrow} = 10^{-4}$ [see Eq.~(\ref{delta_barrier})] for the black full lines, and $\mathcal{D}_{0\uparrow,\downarrow} = 0.1$ for the red dashed ones.}
\label{Transport}
\end{figure*}
	
\subsubsection{Transport}
Let us now consider how the interplay between the YSR impurity bands and ABS manifests itself in electric transport. In order to demonstrate it, we consider two regimes: (1) when broadening of the ABS is governed by magnetic impurities, and (2) when it is dominated by the interface transparency. For simplicity we assume that impurities in the bulk of the superconductor have randomly oriented magnetic moments.

In case (1), when the ABS width is controlled by magnetic impurities, we can still observe both the YSR impurity bands and the ABS, see Fig.~\ref{Transport}(a). At the same time, one has to remember that the ABS are spin-polarized and, as was discussed in the previous section, they repel the YSR bands for two spin-channels in the opposite direction, see Fig.~\ref{Dos_v_x}(b). It is remarkable that in this case, even though there are no spin currents possible in the bulk (impurity spins are oriented randomly), close to the NS surface one can observe a non-zero spin conductance, see Fig.~\ref{Transport}(b). Moreover, as can be seen from Fig.~\ref{Transport}(e), the sub-gap spin currents are highly polarized, with $P\simeq 100\%$ when $\mathcal{D}\ll 1$. Due to low transparency of the interface, sub-gap currents are mostly carried by single-particle excitations via the ABS and YSR bands (and Cooper pairs in between them), see solid line in Fig.~\ref{Transport}(f), and possess negligible noise, see Fig.~\ref{Transport}(c).

For case (2), when broadening of the ABS is governed by the interface transparency, tails of the ABS completely mask the YSR impurity bands, see Fig.~\ref{Transport}(d), but the sub-gap structure can still be seen in the spin conductance, see Fig.~\ref{Transport}(b). Increased transparency favors Andreev reflection and drastically changes the statistics of sub-gap current carriers in the system, see Fig.~\ref{Transport}(f). Indeed, above the gap we still observe $F\approx 1$ (except for the BCS coherence peak at the gap edge), characteristic of single-particle excitations, but the sub-gap transport is governed by two-particle (Andreev) processes, in contrast to the tunneling case considered above, and $F\approx 2$. At voltages probing the ABS, the differential noise is much reduced compared to the charge conductance because the ABS are resonances. On resonance, letting $\Gamma\rightarrow 0$, fluctuations vanish\cite{Lofwander2003} and $F\rightarrow 0$, which is a fingerprint of resonant Andreev reflection. The enhanced role of two-particle Andreev processes substantially reduces the spin-polarization of transferred current, see Fig.~\ref{Transport}(e).

Finally, we point out that the non-zero spin conductance at the interface leads to spin imbalance, an unequal population of quasiparticle branches with opposite spin projections. For randomly oriented impurity spins, spin imbalance at subgap voltages induced by spin-polarized ABS was studied by two of us in a recent publication \cite{Shevtsov_PRB2014}.

\section{Discussion and conclusions}\label{Sec_Concl}
Before summarizing the main findings, let us briefly comment on the applicability of our model. In all the calculations, we have assumed that the YSR states of neighboring impurities have enough wavefunction overlap to form extended metallic impurity bands. The necessary condition for this to happen is the Mott criterion \cite{Mott1990,Balatsky1995} $n_{\mathrm{min}}^{1/3}\xi_0 = 0.2$, which estimates the minimal impurity density necessary for the delocalization transition. In terms of the parameters of our model we obtain $\Gamma_{\mathrm{min}}/2\pi k_BT_{c0} = 0.5(k_BT_c/E_F)^2T_c/T_{c0}$ ($T_{c0}$ is the bulk critical temperature of the superconductor in absence of impurities), meaning that this condition is satisfied for all the values of $\Gamma$ used in our calculations since $k_BT_c\ll E_F$.

In conclusion, we have studied charge and spin transport in NS junctions with a finite density of magnetic impurities. The latter were described within the non-crossing $t$-matrix approach. Considering the two extreme cases -- completely unpolarized and polarized impurities -- we have investigated both electric current and noise across the NS surface. The results of our calculations indicate that adding magnetic impurities to conventional $s$-wave spin-singlet superconductors can substantially extend functionalities of superconducting hybrid devices. We find that in the case when magnetic impurities are ferromagnetically ordered, one can achieve spin-polarizations of the tunneling currents reaching $P\simeq 100\%$. Moreover, for a suitable choice of system parameters, the sign of spin-polarization can be chosen by simply tuning the applied bias. We also demonstrate that even when magnetic impurities are completely unpolarized, one can still inject almost entirely spin-polarized currents across the NS junction. In order to achieve that, we propose to engineer spin-active interfaces between the superconductor and the normal-metal probe. The latter is not just a theoretical model, but was successfully realized in practice by several experimental groups \cite{Khaire_PRL2010,Hubler_PRL2012ABS}. All these features make impurity engineering in superconductors a promising route towards hybrid superconducting spintronic applications. Finally, by analyzing the noise properties of the tunneling currents, we were able to decipher the relative role of single-particle excitations and Cooper pairs, which can be tuned by changing the NS interface transparency. In the tunneling limit, the currents are predominantly carried by single-particle excitations, while Cooper pairs dominate electric transport at high tranparencies.

\begin{acknowledgments}
We would like to acknowledge financial support from the Swedish Research Council. The research of OS was partly supported by the National Science Foundation (Grant DMR-1508730).
\end{acknowledgments}

\appendix

\section{Expressions for $\mathcal{S}^{R-A}$ and $\mathcal{S}^{K}$}\label{App1}
In this Appendix we provide explicit expressions for the two components of the noise formula, Eq.~(\ref{noise_eq}), from the main text. 
We do not give any derivation of these formulas, but rather mention that they originate from a lengthy but straightforward generalization of the procedure described in Ref.~[\onlinecite{Lofwander2003}]. All the expressions presented below are written in terms of the elementary scattering amplitudes given in Table~\ref{scatter_amplitudes2} (see also Fig.~\ref{Fig_AS}).
The spectral part of noise can be written as follows,
\begin{align}
	\mathcal{S}^{R-A} = 4 &+ \left\{r^R_{he},r^A_{he}\right\}-\left\{r^R_{ee},r^A_{ee}\right\}\notag\\
&+\left\{r^R_{eh},r^A_{eh}\right\}-\left\{r^R_{hh},r^A_{hh}\right\},
\end{align}
where $\{a,b\}=ab+ba$ is a regular anti-commutator. On the other hand, the Keldysh component is given by
\begin{align}
	\mathcal{S}^{K} &= s_1 + s_2 + s_3 + s_4 + s_5 + s_6 \nonumber\\
	&+ \tilde{s}_1 + \tilde{s}_2 + \tilde{s}_3 + \tilde{s}_4.
\end{align}
The terms constituting $\mathcal{S}^K$ above can be written as
\begin{widetext} 
\begin{align}
s_1&=
	\left(1+\left\{r^R_{he},r^A_{he}\right\}-\left\{r^R_{ee},r^A_{ee}\right\}-\left\{r^R_{ee}r^A_{he},r^R_{he}r^A_{ee}\right\}
	+r^R_{he}r^A_{he}r^R_{he}r^A_{he}+r^R_{ee}r^A_{ee}r^R_{ee}r^A_{ee}\right)x_N^2,\\
s_2&=
	\overline{t}^R_{he}x_S\overline{t}^A_{he}\overline{t}^R_{he}x_S\overline{t}^A_{he}+\overline{t}^R_{ee}x_S\overline{t}^A_{ee}
	\overline{t}^R_{ee}x_S\overline{t}^A_{ee}-\left\{\overline{t}^R_{ee}x_S\overline{t}^A_{he},\overline{t}^R_{he}x_S\overline{t}^A_{ee}\right\},\\
s_3&=
	\left(\left\{r^R_{he}r^A_{he},\overline{t}^R_{he}x_S\overline{t}^A_{he}\right\}+\left\{r^R_{ee}r^A_{ee},\overline{t}^R_{ee}x_S
	\overline{t}^A_{ee}\right\}-\left\{r^R_{he}r^A_{ee},\overline{t}^R_{ee}x_S\overline{t}^A_{he}\right\}-\left\{r^R_{ee}r^A_{he},
	\overline{t}^R_{he}x_S\overline{t}^A_{ee}\right\}\right)x_N,\\
s_4&=
	\left(\left\{r^R_{he}r^A_{ee},\overline{t}^R_{eh}\tilde{x}_S\overline{t}^A_{hh}\right\}+\left\{r^R_{ee}r^A_{he},\overline{t}^R_{hh}
	\tilde{x}_S\overline{t}^A_{eh}\right\}-\left\{r^R_{ee}r^A_{ee},\overline{t}^R_{eh}\tilde{x}_S\overline{t}^A_{eh}\right\}-
	\left\{r^R_{he}r^A_{he},\overline{t}^R_{hh}\tilde{x}_S\overline{t}^A_{hh}\right\}\right)x_N,\\
s_5&=
	\left(\left\{r^R_{he}r^A_{ee},r^R_{eh}r^A_{hh}\right\}+\left\{r^R_{ee}r^A_{he},r^R_{hh}r^A_{eh}\right\}
	-\left\{r^R_{ee}r^A_{ee},r^R_{eh}r^A_{eh}\right\}-\left\{r^R_{he}r^A_{he},r^R_{hh}r^A_{hh}\right\}\right)x_N\tilde{x}_N,\\
s_6&=
	\left\{\overline{t}^R_{he}x_S\overline{t}^A_{ee},\overline{t}^R_{eh}\tilde{x}_S\overline{t}^A_{hh}\right\}+
	\left\{\overline{t}^R_{ee}x_S\overline{t}^A_{he},\overline{t}^R_{hh}\tilde{x}_S\overline{t}^A_{eh}\right\}-
	\left\{\overline{t}^R_{ee}x_S\overline{t}^A_{ee},\overline{t}^R_{eh}\tilde{x}_S\overline{t}^A_{eh}\right\}-
	\left\{\overline{t}^R_{he}x_S\overline{t}^A_{he},\overline{t}^R_{hh}\tilde{x}_S\overline{t}^A_{hh}\right\}.
	\end{align}
\end{widetext}
Note that the ``tilded" terms are obtained by simply using Eq.~(\ref{tilde_def}). Finally, these formulas reduce to the ones obtained in Ref.~[\onlinecite{Lofwander2003}], if there is no spin dependence in the problem.

%

\end{document}